\documentclass[conference]{IEEEtran}
\IEEEoverridecommandlockouts
\usepackage{cite}
\usepackage{amsmath,amssymb,amsfonts}
\usepackage{amsthm}
\usepackage{algorithm,algorithmic}

\usepackage{graphicx}
\usepackage{textcomp}
\usepackage{xcolor}

\usepackage{subcaption}
\usepackage{comment}
\usepackage{paralist}

\newcommand{\myminer}{{\sf FlexHMiner}}

\newcommand{\mytitle}{Discovering Hierarchical Processes \\ Using Flexible Activity Trees for Event Abstraction}

\usepackage{hyperref}       

\usepackage{scrextend} 

\usepackage{booktabs}
\usepackage{multicol}
\usepackage{multirow}
\usepackage{array}
\newcolumntype{L}[1]{>{\raggedright\let\newline\\\arraybackslash\hspace{0pt}}m{#1}}
\newcolumntype{C}[1]{>{\centering\let\newline\\\arraybackslash\hspace{0pt}}m{#1}}
\newcolumntype{R}[1]{>{\raggedleft\let\newline\\\arraybackslash\hspace{0pt}}m{#1}}

\newcommand{\columnn}[1]{\emph{#1}}

\newcommand{\todo}[1]{}
\newcommand{\term}[1]{\emph{#1}}
\theoremstyle{definition}
\newtheorem{definition}{Definition}
\newtheorem{example}{Example}

\newcommand{\abs}[1]{\ensuremath{\mid #1 \mid}}
\newcommand{\mi}[1]{\ensuremath{\mathit{#1}}}

\newcommand{\floor}[1]{\lfloor #1 \rfloor}

\newcommand{\preset}[1]{\ensuremath{{{^\bullet}#1}}}
\newcommand{\postset}[1]{\ensuremath{#1^\bullet}}

\usepackage{xcolor}
\definecolor{mygray}{gray}{0.6}
\usepackage{algorithm,algorithmic}

\renewcommand{\algorithmiccomment}[1]{\quad\textcolor{mygray}{\{\textit{#1}\}}}

\newcommand{\actlabel}[1]{\text{\texttt{{#1}}}}

\newcommand{\chld}{\ensuremath{\gamma}}

\newcommand{\sylogacts}{\ensuremath{\Sigma}}
\newcommand{\syallacts}{\ensuremath{A}}

\newcommand{\sydalg}{\ensuremath{D}}
\newcommand{\symaplog}{\ensuremath{\alpha}}
\newcommand{\symapmodel}{\ensuremath{\beta}}

\usepackage{xspace}
\newcommand{\aam}{FlexHMiner\xspace}
\newcommand{\aamdk}{DK-FH\xspace}
\newcommand{\aamr}{RC-FH\xspace}
\newcommand{\aamflat}{F-FH\xspace}
\newcommand{\disflat}{F*\xspace}

\newcommand{\lowl}[2]{\ensuremath{f_\downarrow({#1}, {#2})}}
\newcommand{\highl}[2]{\ensuremath{f_\uparrow({#1}, {#2})}}

\newcommand{\qfit}{\ensuremath{\mi{fi}}}
\newcommand{\qprec}{\ensuremath{\mi{pr}}}

\def\BibTeX{{\rm B\kern-.05em{\sc i\kern-.025em b}\kern-.08em
		T\kern-.1667em\lower.7ex\hbox{E}\kern-.125emX}}

\begin{document}

\title{\mytitle\\
}

	
	\author{
		\IEEEauthorblockN{
			Xixi Lu\IEEEauthorrefmark{1},
			Avigdor Gal\IEEEauthorrefmark{2}, and 
		    Hajo A. Reijers\IEEEauthorrefmark{1}}
		\IEEEauthorblockA{\IEEEauthorrefmark{1}%
			\textit{Dept. of Information and Computing Sciences} \\
			\textit{Utrecht University}, Utrecht, The Netherlands \\
			x.lu@uu.nl, h.a.reijers@uu.nl} 
				\IEEEauthorblockA{
					\IEEEauthorrefmark{2}%
						\textit{Faculty of Industrial Engineering and Management} \\
					\textit{Technion -- Israel Institute of Technology}, Haifa, Israel \\
					avigal@technion.ac.il}
			}
%

\maketitle

\begin{abstract}
%
Processes, such as patient pathways, can be very complex, comprising of hundreds of activities and dozens of interleaved subprocesses.
While existing process discovery algorithms have proven to construct models of high quality on clean logs of structured processes, it still remains a challenge when the algorithms are being applied to logs of complex processes. 
The creation of a multi-level, hierarchical representation of a process can help to manage this complexity. However, current approaches that pursue this idea suffer from a variety of weaknesses. In particular, they do not deal well with interleaving subprocesses.
In this paper, we propose \myminer, a three-step approach to discover processes with multi-level interleaved subprocesses. 
We implemented \myminer~in the open source Process Mining toolkit ProM. We used seven real-life logs to compare the qualities of hierarchical models discovered using domain knowledge, random clustering, and flat approaches.  Our results indicate that the hierarchical process models that the \myminer~generates compare favorably to approaches that do not exploit hierarchy.

%




\end{abstract}


\section{Introduction}\label{sec:intro}


Complex processes often comprise of hundreds of activities and dozens of subprocesses that run in parallel~\cite{DBLP:journals/sosym/LeemansFA18,DBLP:conf/bpm/LuTHR19}. 
For example, in healthcare, a patient follows a certain process that consists of several procedures (e.g., lab test and surgery) for treating a medical condition. 
%
Studies have shown qualitatively that using hierarchical process models to represent complex processes can improve understandability and simplicity by ``hiding less relevant information''~\cite{DBLP:journals/is/ReijersMD11}. In particular, the use of modularized, hierarchical process models, where process activities are collected into subprocesses, can be presented in a condensed higher-level presentation, revealing more details upon request.


\term{Process discovery}, a prominent task of process mining, aims at automatically constructing a process model from an event log. Over the years, dozens of discovery algorithms have been proposed~\cite{DBLP:conf/apn/LeemansFA14,DBLP:journals/tkde/CarmonaC14,DBLP:conf/cidm/WeijtersR11,DBLP:conf/icdm/AugustoCDR17}. These have proven to perform very well on relatively clean logs of structured processes. However, given the complexity of real-life processes, these algorithms tend to discover overfitted or overgeneralized models that are difficult to comprehend~\cite{DBLP:journals/tkde/AugustoCDRMMMS19}. Existing discovery algorithms tend to disregard any hierarchical decomposition, whereas discovering hierarchical process models may help decrease the complexity of the discovery task. 

\begin{figure}[tb]
	\begin{subfigure}{.23\textwidth}
		\centering
		\includegraphics[height=10cm]{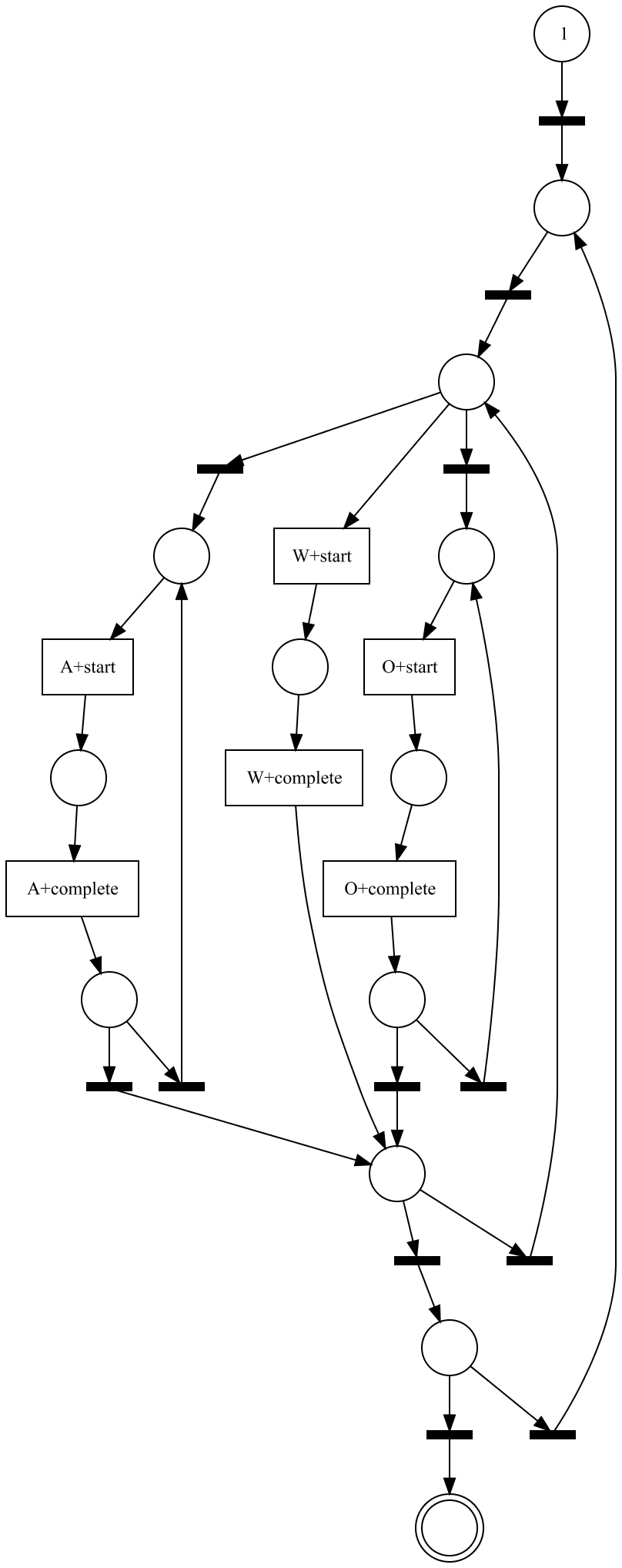}  
		\caption{The sequential subprocessses discovered by SCM~\cite{DBLP:conf/wcre/LeemansAB18}. }
		\label{fig:rwSCMvsAAMbpi2012highscm}
	\end{subfigure}%
	\hfill
	\begin{subfigure}{.24\textwidth}
		\centering
		\includegraphics[height=10cm]{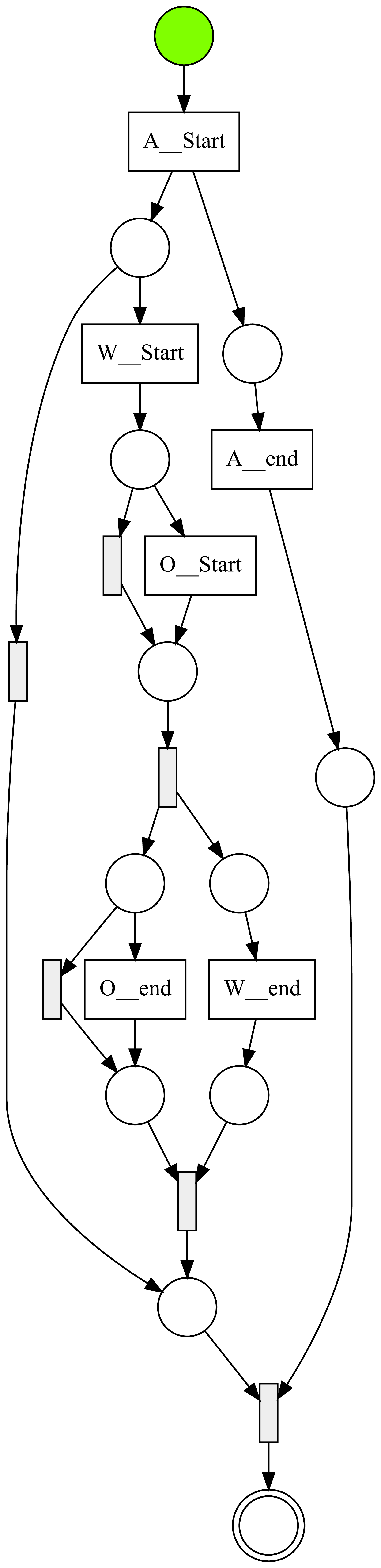}  
		\caption{The interleaving subprocesses discovered by our \aamdk.}
		\label{fig:rwSCMvsAAMbpi2012highaam}
	\end{subfigure}
	\caption{Difference in the discovered root models for the BPIC2012 log.}
	\label{fig:rwSCMvsAAMbpi2012high}
\end{figure}

Few hierarchical process discovery algorithms have been proposed~\cite{DBLP:conf/bpm/BoseA09,DBLP:conf/bpm/MaggiSR14,DBLP:conf/otm/WangWYSW15,DBLP:journals/is/ConfortiDGR16,DBLP:conf/bpm/MannhardtLRAT16,DBLP:conf/wcre/LeemansAB18}. 
%
While some approaches, such as \cite{DBLP:journals/is/ConfortiDGR16},
aim to automatically detect subprocesses, these algorithms are rigid in their assumptions and require extensive knowledge encoding for every event in the log regarding the causalities between the activities.
Other approaches assume that subprocesses do not interleave~\cite{DBLP:conf/bpm/BoseA09,DBLP:conf/wcre/LeemansAB18}, which lead to discovering inaccurately, overly segmented models (see~\autoref{fig:rwSCMvsAAMbpi2012high}). 
It is also unclear how these approaches compare to conventional, flat models in terms of quality measures such as fitness and precision.



In this work, we propose FlexHMiner (FH), a three-step approach for the discovery of hierarchal models. 
We formalize the concept of \emph{activity tree} and \emph{event abstraction}, which allows us to be flexible in the ways of computing the process hierarchy. We illustrate this flexibility by proposing three different techniques to discover an activity tree:
(1)  a fully domain-based approach (\aamdk), 
(2) a random approach (\aamr), and
(3) a fall-back, flat activity tree (\aamflat).  
After obtaining an activity tree, the second step of our approach is to compute the logs for each subprocess using log abstraction and log projection. Finally, \myminer~discovers a subprocess model for each subprocess by leveraging the capabilities of existing discovery algorithms. 
Using the domain-based approach as the gold standard and the flat tree approach as base line, we compare the three ways of discovering an activity tree using seven real-life logs.  


The main contribution of this work is a novel approach to discover hierarchical process models from event logs, which (1) can handle multi-level interleaving subprocesses and (2) is flexible in its data requirements so that it can adapt to the situations when there is domain knowledge and when there is not. 
%
%
%
We evaluated \aam using seven real-life, benchmark event logs\footnote{\scriptsize The source code and the results can be found at: \href{https://github.com/xxlu/prom-FlexHMiner}{github.com/xxlu/prom-FlexHMiner}}. 
In the remainder, we define the research problem in~\autoref{sec:hierarchy}. The proposed approach is described in~\autoref{sec:approach}.
The evaluation results are presented in~\autoref{sec:evaluation}. \autoref{sec:relatedwork} discusses related work, and \autoref{sec:conclusion} concludes the paper .

\section{Activity Tree and Process Hierarchy}
\label{sec:hierarchy}

\begin{table}[tb]
\centering
\caption{An example of an event log of a healthcare process.}
\resizebox{.48\textwidth}{!}{
    \begin{tabular}{rrllll}
    \toprule
    $\mathcal{E}$ & Patient & Description & Subprocess    & Act & Timestamp \\
    \midrule
    $e_1$     & 101  & Visit & Contact & C\_Vi & 10-10-2019 \\
    $e_2$     & 101  & Calcium & Labtest & L\_Ca & 11-10-2019 \\
    $e_3$     & 101  & Register & Contact & C\_Re & 12-10-2019 \\
    $e_4$     & 101  & Glucose & Labtest & L\_Gl & 13-10-2019 \\
    $e_5$     & 101  & Consultation & Contact & C\_Cs & 14-10-2019 \\
    $e_6$     & 101  & Consultation & Contact & C\_Cs & 15-10-2019 \\
    $e_7$     & 102  & Register & Contact & C\_Re & 16-10-2019 \\
    $e_8$     & 102  & Glucose & Labtest & L\_Gl & 17-10-2019 \\
	  ...    & ...  & ... & ... & ... & ...  \\
    \bottomrule
\end{tabular}%
}
\label{tab:eventloglistexample}%
\end{table}%



\begin{figure}[tb]
	\centerline{\includegraphics[width=.5\textwidth]{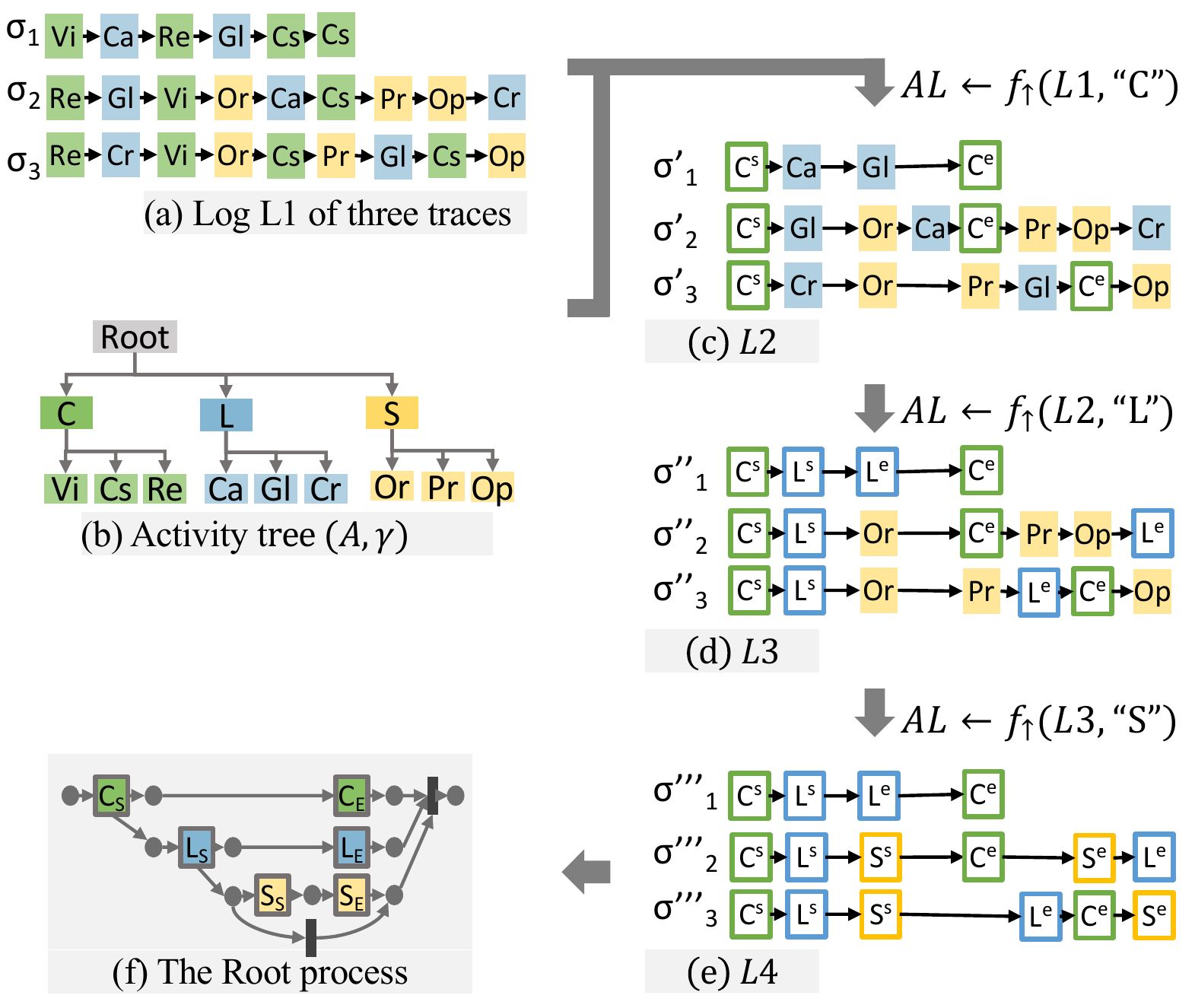}}
	\caption{\aam applied on the running example.}
	\label{fig:exampleApproach}
\end{figure}

\paragraph*{Preliminaries} Let $\sigma  = \langle a_1, \cdots, a_n\rangle \in \sylogacts^*$ be a sequence of activities, which is also called a \emph{trace}. Let a multiset $L \subseteq \mathbf{B}(\sylogacts^*)$ of traces be an \emph{event log}. 
\autoref{tab:eventloglistexample} shows a list of events; \autoref{fig:exampleApproach}(a) shows the sequential traces, in a graphical representation, where each square represents an event that is labeled with an activity.
Given an event log, a process discovery algorithm $D$ automatically constructs a process $M$ (e.g., in Petri net notation). 
The quality of such a model $M$ can assessed with respect to the log $L$ using four quality dimensions: fitness, precision, generalization, and complexity~\cite{DBLP:journals/tkde/AugustoCDRMMMS19}. 

To discover a hierarchical process model, we leverage a rather known concept that represents the process hierarchical information and call it \term{activity tree}. We define an activity tree as the hierarchical relations between activities of a process. Formally, an activity tree is a non-overlapping, hierarchical clustering of activities.


\begin{definition}[Activity tree]
Let $\sylogacts$ be a set of activities and $L$ an \emph{event log} over $\sylogacts$. Let $\syallacts$ be a superset of $\sylogacts$, i.e., $\sylogacts \subset \syallacts$. 
%
%
Function $\chld : \syallacts \rightarrow \mathcal{P}(\syallacts)$ is a mapping that maps each activity $x \in \syallacts$ to a set of activities $X \subset \syallacts$ as the children of $x$. An activity tree $(\syallacts, \chld)$ is valid for $L$, if and only if: 
\begin{enumerate}
\item
The children of any two labels do not overlap, i.e., for each $x, y \in \syallacts$, $x \neq y \Leftrightarrow \chld(x) \cap \chld(y) = \emptyset$.  
\item
The union of the leaves is $\sylogacts$, i.e., the set of activities that occurred in log $L$, i.e., $\{ x \in \syallacts \mid \chld(x) = \emptyset\} = \sylogacts$. 
\item 
The tree $(\syallacts, \chld)$ is connected, i.e., for each $x \in \syallacts$, either there is $y \in \syallacts$ such that $x \in \chld(y)$, or $x$ is the \emph{root}.  
\end{enumerate}
\end{definition}

Note that constraints (2) and (3) imply that for all $x \in \syallacts$, $x \notin \chld(x)$. Furthermore, it is also possible to define an activity tree as a directed acyclic graph where each node represents an activity and has only one incoming edge, except the \emph{root} node, which does not have any incoming edge. 


%
\begin{example}
\autoref{fig:exampleApproach}(b) exemplifies the activity tree $(\syallacts, \chld)$, where $\syallacts = \{\actlabel{root},\allowbreak \actlabel{C}, \allowbreak \actlabel{L},\allowbreak \actlabel{S}, \actlabel{Vi}, \actlabel{Cs}, \actlabel{Re}, \actlabel{Ca}, \actlabel{Gl}, \actlabel{Cr},\allowbreak  \actlabel{Or}, \allowbreak \actlabel{Pr}, \actlabel{Op}\}$, and, for example, $\chld(\actlabel{C}) = \{\actlabel{Vi}, \actlabel{Cs}, \actlabel{Re}\}$. 
\end{example}

We call the node that has no parent in $\chld$ the root node, and the corresponding process the \term{root process}.  For each $a \in \syallacts$, $\chld(a) \neq \emptyset$, we call it a (sub)process $a$ with $\chld(a)$ as its activities. 

We define the height of a node in the activity tree, which is later used to recursively compute the abstracted logs bottom-up. 
Let $height : \syallacts \rightarrow \mathbb{N}^0$ be a function that maps each $x \in \syallacts$ to the height (a non-negative integer) of $x$ in the activity tree $(\syallacts, \chld)$. 
For each $x \in \syallacts$, if $\chld(x) = \emptyset$, then  $height(x) = 0$, else $height(x) = 1 + \mathit{Max}_{c \in \chld(x)} height(c)$. 
A process model with a maximal height of an activity tree is 1, is called {\em flat}.  

\section{The \aam Algorithm}
\label{sec:approach}

In this section, we discuss the three steps of the \aam approach: (1) compute an activity tree, (2) compute abstracted logs, and (3)
 compute subprocess models. 

\subsection{Computing Activity Tree}\label{sec:acthierarchy}
For the first step, computing an activity tree, we present three different methods: one supervised using domain knowledge, one automated using random clustering, and one fall-back using flat tree.

\subsubsection{Using Domain Knowledge}
Domain knowledge can be used, as a gold standard, to create an activity tree. This can be done manually. In other situations, the log itself may already contain an encoding of human knowledge that can be utilized in creating the activity tree. In \autoref{tab:eventloglistexample}, the column \columnn{Act} demonstrates such an encoding of hierarchy. For instance, the activity label \actlabel{C\_Vi}, indicating that the event belongs to subprocess $\actlabel{C}$. Following the same strategy as the State Chart Miner (SCM)~\cite{DBLP:conf/wcre/LeemansAB18}, a simple text parser is built to split the labels of activities and to convert the domain knowledge into an activity tree.   
For our experiment, we found seven real-life logs of two processes, where such information regarding the hierarchy is readily encoded in the activity labels
\footnote{\label{foo:bpic12}See the description of the BPI Challenge 2012 at \url{https://www.win.tue.nl/bpi/doku.php?id=2012:challenge}: \textit{``The event log is a merger of three intertwined sub processes. The first letter of each task name identifies from which sub process (source) it originated.''}} 
\footnote{\label{foo:bpic15}See the description of the BPI Challenge 2015 at \url{https://doi.org/10.4121/uuid:31a308ef-c844-48da-948c-305d167a0ec1}.: \emph{``The first two digits as well as the characters} [of the activity label] \emph{indicate the subprocess the activity belongs to. For instance ... \actlabel{01\_BB\_xxx} indicates the `objections and complaints' (`Beroep en Bezwaar' in Dutch) subprocess.''}}.
%
%

\subsubsection{Using Random Clustering}  Let $L \subset \mathbf{B}(\sylogacts^*)$ be an event log. Let $\mi{maxSize}  \geq 2$ be the indicated maximal size of subprocesses.  The random clustering algorithm, as listed in Algorithm~\autoref{alg:activitytree}, takes $\sylogacts$ and $\mi{maxSize}$ as inputs and creates random subprocesses of size less than $\mi{maxSize}$. It first initializes the activity tree $(\syallacts, \chld)$ using $\sylogacts$ (see Line 1). It then uses $\mi{maxSize}$ to determine the number $n$ of subprocesses at the current height (see Line 2-5). Next, it creates $n$ parent nodes as $P$ and simply assigns each activity $c \in C$ randomly to a cluster $p \in P$ (see Line 6-12), while updating $\syallacts$ and $\chld$ accordingly (see Line 8 and 11). To decide whether to continue with the next height, the current set of parent nodes $P$ becomes the set of child nodes $C$ (see Line 13): if the size of $C$ (i.e., $\abs{C}$) is greater than $\mi{maxSize}$, then the algorithm creates another level of parent nodes (as abstracted processes); otherwise, the \emph{while} loop terminates, and the root node is added (see Line 15 and 16). In this paper, $\mi{maxSize} $ is set to 10 as default. We use the random clustering to show an unsupervised way to create an \emph{activity tree} and to investigate a possible baseline for the qualities of hierarchical models. 

\begin{algorithm}[b]
	\caption{Compute random tree $(\syallacts, \chld)$ }
	\label{alg:activitytree}
	\algsetup{linenosize=\small}
	\scriptsize
	\begin{algorithmic}[1]
		\REQUIRE the input log $L$ over  $\sylogacts$, and the maximal size of a process $\mi{maxSize}$
		\ENSURE activity tree $(\syallacts, \chld)$, 
		\STATE $\syallacts \leftarrow \sylogacts$; and {\bf for} $a \in \sylogacts$ {\bf do} $\chld(a) \leftarrow \emptyset$
		\algorithmiccomment{initiate $\syallacts$ and the leaves of $\chld$}
		\STATE $C \leftarrow \sylogacts$	\algorithmiccomment{Use $C$ to represent the child activities at the current height}

		\WHILE{ $\abs{C} > \mi{maxSize}$} 
		\STATE \algorithmiccomment{create parent clusters $P$ to ensure the size of a process is at most $\mi{maxSize}$, and apply clustering to assign each $c \in C$ to a parent cluster $p$}
		\STATE $\mi{n} \leftarrow   \floor{(\abs{C}-1) / \mi{maxSize}} + 1$ \algorithmiccomment{Calculate the number of parents}
		\STATE  Create labels $p_1, \cdots, p_n$ \algorithmiccomment{Create $n$ random parent nodes/processes}
		\STATE Initiate $P \leftarrow \{p_1, \cdots, p_n\}$:  {\bf for} $p \in P$ {\bf do} $\chld(p) \leftarrow \emptyset$
		\STATE $\syallacts \leftarrow \syallacts \cup P$
		\FOR{ $c \in C$}
		\STATE Select a random $p \in P$, where $\abs{\chld(p)} < \mi{maxSize}$
		\STATE $\chld(p) \leftarrow \chld(p) \cup \{c\}$
		\ENDFOR	\algorithmiccomment{all children have been assigned to a parent process}
		\STATE $C \leftarrow P$
		\ENDWHILE \algorithmiccomment{$\abs{C} \leq \mi{maxSize}$}
		\STATE $\syallacts \leftarrow \syallacts \cup \{root\}$ \algorithmiccomment{Create the root node/process and add to $\syallacts$}
		\STATE {\bf for} $c \in C$ {\bf do} $\chld(root) \leftarrow \chld(root) \cup \{c\}$ \algorithmiccomment{Assign $c \in C$ to the root process}
		\RETURN $(\syallacts, \chld)$
	\end{algorithmic}
\end{algorithm}

\subsubsection{Using Flat Tree} Given a log $L$ over activities $\sylogacts$, a flat activity tree $(\syallacts, \chld)$ (with height 1) is constructed as follows. We have $\syallacts = \sylogacts \cup \{root\}$. For all $a \in \sylogacts$, $\chld(a) = \emptyset$; and for $root \in \syallacts$, $\chld(root) = \sylogacts$. We use the flat tree approach as a fall-back.

\subsection{Computing Abstracted Logs and Models}
Given an activity tree $(\syallacts, \chld)$ and a log $L$, the second step uses the projection ($f_\downarrow$) and abstraction ($f_\uparrow$) functions to recursively compute sublogs for each non-leaf node in the tree. We first discuss the projection and abstraction functions, after which we present the algorithm. 

\subsubsection{Projection}

Given a log $L$, an activity tree $(\syallacts, \chld)$, and a subprocess $sp \in \syallacts$, the projection of $L$ on the activities $\chld(sp)$ of $sp$ is rather straightforward and standard, which allows us to create a corresponding log for $sp$. The projection function simply retains the events of activities $\chld(sp)$ and remove the rest. 
%
%
Formally, given a trace $\sigma = \langle e \rangle \cdot \sigma'$: if $e \in \chld(sp)$, $\lowl{\sigma}{sp} =  \langle e \rangle \cdot \lowl{\sigma'}{sp}$, otherwise, $\lowl{\sigma}{sp} =  \langle \rangle \cdot \lowl{\sigma'}{sp}$. We overload the function for any log $L$:
$\lowl{L}{sp} = \bigcup_{\sigma \in L} \lowl{\sigma}{sp}$, if $\lowl{\sigma}{sp} \neq \langle \rangle$. 
%
\begin{example}
For example,  given the (simplified) trace $\sigma_1 = \langle \actlabel{Vi}, \actlabel{Ca}, \actlabel{Re}, \actlabel{Gl}, \actlabel{Cs}, \actlabel{Cs}\rangle$,
the subprocess $\actlabel{C}$, and the activity-hierarchy $\chld$ where $\chld(\actlabel{C}) = \{\actlabel{Vi}, \actlabel{Cs}, \actlabel{Re}\}$ (as shown in~\autoref{fig:exampleApproach}(a) and (b)) then the trace of \actlabel{C} is computed as 
$\lowl{\sigma_1}{\actlabel{C}} = \langle \actlabel{Vi}, \actlabel{Re}, \actlabel{Cs}, \actlabel{Cs} \rangle$.
\end{example}

\subsubsection{Abstraction}
The \term{abstraction function} $f_\uparrow$ returns the abstracted, intermediate log after abstracting (removing) the internal behavior of a subprocess $sp$. Essentially, the abstraction function hides irrelevant internal behavior by not retaining the detailed events of a subprocess; it only keeps the \emph{relevant behavior}. 
In this paper, we consider both the \emph{start} and the \emph{end events} of a subprocess as the relevant behavior. In this way, the abstracted log only records when a subprocess starts and ends\footnote{\scriptsize Note that other abstraction functions can be conceived, e.g., a function that will only render the start or the end event of the subprocess.}. If a subprocess $\actlabel{x}$ has different start activities (e.g., $\actlabel{Vi}$ and $\actlabel{Re}$), they are abstracted into a single start activity $\actlabel{x}^s$ (e.g., $\actlabel{C}^s$, see~\autoref{fig:exampleApproach}(c)). The same holds for the end activities, which are abstracted into a single end activity $\actlabel{x}^e$. 
%

Let $\mi{SP} = \chld(sp)$ denote the set of activities of subprocess $sp$. 
Let $I_s$ denote the index of the first event of $sp$ in $\sigma$ as $I_s = \min_{e \in \sigma \wedge e \in \mi{SP}} \sigma[e]$. Similarly, we use $I_e$ to refer to the index of the last event of $sp$ in $\sigma$, i.e.,  $I_e = \max_{e \in \sigma \wedge e \in \mi{SP}} \sigma[e]$. We define $f_\uparrow$ as follows:

{\scriptsize
\begin{align}
\highl{\sigma}{\mi{SP}} &= 
\begin{cases}
\langle e \rangle \cdot  \highl{\sigma'}{\mi{SP}} & e \notin \mi{SP}  \\
\langle sp^s\rangle \cdot  \highl{\sigma'}{\mi{SP}} &  e \in \mi{SP}, \sigma[e] = I_s\\
\langle sp^e \rangle \cdot  \highl{\sigma'}{\mi{SP}} & e \in \mi{SP}, \sigma[e] = I_e\\
\langle \rangle \cdot  \highl{\sigma'}{\mi{SP}}  & e \in \mi{SP} \\ & \quad \wedge \quad I_s < \sigma[e] < I_e
\end{cases}
\end{align}
}

We overload the function for any log $L$, i.e., $\highl{L}{sp} = \bigcup_{\sigma \in L} \highl{\sigma}{sp}$
\footnote{\scriptsize 
	For the sake of brevity, we consider the labels $a = a^s = a^e$ for both projection and abstraction functions and only distinguish them when we apply a discovery algorithm.}.

\begin{example}\label{exa:abstraction}
	For example, after abstracting subprocess \actlabel{C} from $\sigma_1$, we have $\highl{\sigma_1}{\actlabel{C}}   = \langle \actlabel{C}^{s}, \actlabel{Ca}, \actlabel{Gl},\allowbreak \actlabel{C}^{e}\rangle$. Similar, the trace after abstracting subprocess \actlabel{L} is $\highl{\sigma_1}{\actlabel{L}}   = \langle \actlabel{Vi}, \actlabel{L}^{s}, \actlabel{Re}, \actlabel{L}^{e}, \actlabel{Cs}, \actlabel{Cs} \rangle$. To obtain the parent-trace, we apply $f_\uparrow$ recursively; for instance, $\highl{\highl{\sigma_1}{\actlabel{C}}}{\actlabel{L}} = \highl{\highl{\sigma_1}{\actlabel{L}}}{\actlabel{C}} =   \langle \actlabel{C}^s, \actlabel{L}^s, \actlabel{L}^e, \actlabel{C}^e \rangle = \sigma''_1$, see~\autoref{fig:exampleApproach}(d).
\end{example}

%
%

%
%

The abstraction function $f_\uparrow$ is commutative and associative for \emph{non-related} subprocesses\footnote{\scriptsize Two nodes are non-related if they do not share ancestors or descendants in the activity tree)}. 
These two properties allow Algorithm~\ref{alg:loghierarchy} to use the abstraction function in a recursive, bottom-up manner and to go iteratively through the nodes. The algorithm creates an abstracted log at each height of the activity tree, which is discussed in the next section.

\subsubsection{Recursion}

The algorithm computes the log mapping $\symaplog$ and the model mapping $\symapmodel$ with three inputs: (1) an event log $L$, (2) an activity tree $(A, \chld)$, and (3) a flat process discovery algorithm $D$. 
It applies the projection function bottom-up for each subprocess on the abstracted log $\mi{AL}$, starting with height~1 until but does not include the \emph{root} process (see Lines 2 - 13). For every subprocess $p \in P$ of the same height, the algorithm applies the projection function to obtain the sublog $L_p$ for $p$ (Line 8) and updates the log mapping and model mapping (see Lines 9 and 10). The algorithm then computes the intermediate, abstracted log $\mi{AL}$ by using the abstraction function $f_\uparrow$ (see Line 11). The abstracted log $AL$ is updated after each $p \in P$ (see Lines 5 - 12) at height $i$. The algorithm terminates once it has completed processing the root process. It returns a \emph{hierarchical model} $(\syallacts, \chld, \alpha, \beta)$, where the log mapping $\alpha$ maps each (sub)process $p \in \syallacts$, $\chld(p) \neq \emptyset$, to the associated log $L_p$, and the model mapping $\beta$ maps each $p$ to the associated model $M_p = D(L_p)$ of (sub)process $p$. 

\begin{example}
Given the log $L1$ and activity tree $(\syallacts, \chld)$), respectively, shown in \autoref{fig:exampleApproach}(a)  and (b), the three subprocesses \actlabel{C},  \actlabel{L}, and \actlabel{S} have height 1.  \autoref{fig:exampleApproach}(c), (d), and (e) show the abstracted log $AL$ obtained after each iteration of the inner-loop at Lines 5-11. For example, in the first iteration (see \autoref{fig:exampleApproach} (c)), applying $\highl{L1}{\actlabel{C}}$ on log $L1$, we obtain $AL = L2$ where the activities of \actlabel{C} is removed from $L2$ and only the start and the end of \actlabel{C} are retained (as discussed in Example \autoref{exa:abstraction}). In the second iteration, the algorithm continues with log $AL$ and applies $\highl{AL}{\actlabel{L}}$, see \autoref{fig:exampleApproach} (d). 
\end{example}


 \begin{algorithm}[b]
 \caption{Compute log hierarchy $\symaplog$ and models $\symapmodel$}
 \label{alg:loghierarchy}
 \algsetup{linenosize=\small}
 \scriptsize
 \begin{algorithmic}[1]
 \REQUIRE Log $L$, activity tree $(\syallacts, \chld)$, and discovery algorithm $\sydalg$
 \ENSURE Log mapping $\symaplog$ and model mapping $\symapmodel$
 \STATE $\mi{maxHeight} \leftarrow \mathit{Max}_{c \in \syallacts} height_\chld(c)$ \algorithmiccomment{calculate the maximal height of the tree}
 \STATE $\mi{AL} \leftarrow L $
 \FOR{$i = 1$ to $\mi{maxHeight - 1}$}
 \STATE $\mi{P} \leftarrow \{ p \in \syallacts \mid \mi{height}(p) = i \}$
 \FOR{$p \in \mi{P} $} 
 \STATE \algorithmiccomment{Iteratively go through each $p\in P$ at the same height $i$ and perform log projection and abstraction}\\
	 \STATE $C \leftarrow \chld(p)$ \algorithmiccomment{get the activities $C$ of subprocess $p$}
	 \STATE $L_{p} \leftarrow \lowl{\mi{AL}}{C} $  
	 \algorithmiccomment{project the log so far on the activities $C$}
	 
	 \STATE $\symaplog(p) \leftarrow L_{p}$
	 \STATE $\symapmodel(p) \leftarrow D(L_{p})$
	 
	 \STATE $\mi{AL} \leftarrow \highl{\mi{AL}}{p} $ 
	 \algorithmiccomment{abstract the log so far using the activities $C$ }	 
 \ENDFOR \algorithmiccomment{the log so far $AL$ has been abstracted from $P$ to height $i$ }	 
 \ENDFOR \algorithmiccomment{$i == \mi{maxHeight}$}	 
 \STATE $\symaplog(root) \leftarrow \mi{AL}$  \algorithmiccomment{map the root to the resulted abstracted log}	 
 \STATE $\symapmodel(root) \leftarrow D(\mi{AL})$ 
 \algorithmiccomment{map the root to the discoverede model}	 
 \RETURN $(\syallacts, \chld, \symaplog, \symapmodel)$
 \end{algorithmic}
 \end{algorithm}

\section{Empirical Evaluation}
\label{sec:evaluation}
We implemented the \aam approach in the process mining toolkit ProM\footnote{\scriptsize\url{http://www.promtools.org/}. The source code and results: \href{https://github.com/xxlu/prom-FlexHMiner}{github.com/xxlu/prom-FlexHMiner}}. We evaluated the quality of the models created by \aam using seven real-life data sets and compared the results.

In the following, we discuss the data sets and the experimental setup, followed by our empirical analysis.
%
All experiments are run on an Intel Core i7-
8550U 1.80GHZ with a processing unit of a 16 GB HP-Elitebook running Windows 10 Enterprise. 

\begin{table}[tbp]
	\caption{Statistical information of the event logs.}
	\begin{center}
		\resizebox{0.5\textwidth}{!}{
			 \begin{tabular}{|l|r|r|r|r|r|r|}
			 \toprule
			 \textbf{Data} & \textbf{\#acts} & \textbf{\#evts} & \textbf{\#case} & \textbf{\#dpi} & \textbf{avg e/c} & \textbf{max e/c} \\
			 \midrule
			 BPIC12 * & 36    & 262,200 & 13,087 & 4,366 & 20    & 175 \\
			 BPIC17f * & 18    & 337,995 & 21,861 & 1,024 & 18    & 32 \\
			 BPIC15\_1f * & 70    & 21,656 & 902   & 295   & 24    & 50 \\
			 BPIC15\_2f *& 82    & 24,678 & 681   & 420   & 36    & 63 \\
			 BPIC15\_3f *& 62    & 43,786 & 1,369 & 826   & 32    & 54 \\
			 BPIC15\_4f *& 65    & 29,403 & 860   & 451   & 34    & 54 \\
			 BPIC15\_5f *& 74    & 30,030 & 975   & 446   & 31    & 61 \\
			 \bottomrule
			 \end{tabular}%

		}
		\label{tab:logs}
	\end{center}
\end{table}



\subsection{Data sets}\label{sec:data}
An overview of the statistical information, including the number of distinct activities (acts), events (evts), cases, distinct sequences (dpi), and of events per case (e/c) of the seven logs, is given in~\autoref{tab:logs}. 
%
These logs are the filtered logs used in~\cite{DBLP:journals/tkde/AugustoCDRMMMS19} as a benchmark\footnote{\scriptsize\url{https://doi.org/10.4121/uuid:adc42403-9a38-48dc-9f0a-a0a49bfb6371}}, which allows us to compare our result with the qualities of the flat models reported in~\cite{DBLP:journals/tkde/AugustoCDRMMMS19}.  

\subsection{Experimental Setups}\label{sec:experimentsetup}
For computing activity tree, we used the three methods: random (\aamr), flat tree (\aamflat), and domain knowledge (\aamdk). 
For computing hierarchical models, we use two state-of-the-art process discovery algorithms as $D$, namely the Inductive Miner with 0.2 path filtering (IMf) and the Split Miner (SM)~\cite{DBLP:conf/icdm/AugustoCDR17} with standard parameter settings. According to the recent work of Augusto et al.~\cite{DBLP:journals/tkde/AugustoCDRMMMS19}, these two algorithms outperform others in terms of the four quality measures and execution time. 
%

We run these six configurations on the seven data sets. For each log and for each of the two flat discovery algorithms, we also report the results in ~\cite{DBLP:journals/tkde/AugustoCDRMMMS19}, for which we use \disflat to represent. Thus, in total eight rows for each data set. 


\subsection{Model Quality Measures}
\label{sec:evalmeasures}

We assess the quality of a model $M$, with respect to a log $L$, using the following four dimensions:
\begin{itemize}
	\item
For \emph{fitness} $\qfit(L,M) \in [0, 1]$, we use the alignment based fitness defined by Adriansyah et al.~\cite{adriansyah2011conformance}, also used in~\cite{DBLP:journals/tkde/AugustoCDRMMMS19}. 

\item
For \emph{precision} ($\qprec(L,M) \in [0, 1]$), we use the measure defined in~\cite{DBLP:conf/bpm/MunozGamaC10}, known as ETC-align.

\item
We also compute the F1-score, which is the harmonic mean of fitness and precision:
$f1 (M, L) = 2 * \frac{\qfit(L,M) * \qprec(L,M)}{\qfit(L,M) + \qprec(L,M)}$.

\item
For \emph{generalization}, we follow again the same approach adopted in~\cite{DBLP:journals/tkde/AugustoCDRMMMS19}. 
We divide the log into $k = 3$ subsets: 
$L$ is randomly divided into $L_1$, $L_2$, $L_3$, $ge(L, M) = \frac{1}{3}  \sum_{1 \leq i \leq 3} 2 * \frac{fi(L_i, M_i)  * pr(L, M_i)}{fi(L_i, M_i)  + pr(L, M_i)}$ where $M_i$ corresponds to $L_i$.

\item
Finally, for \emph{complexity}, we report the \emph{size} (number of nodes) and the Control-Flow Complexity (CFC) of a model~~\cite{DBLP:journals/tkde/AugustoCDRMMMS19}.  Let $PN = (P, T, F, l, m_i, m_f)$ be a Petri Net. $\mi{Size}(PN) = \abs{P} + \abs{T}$.
%

{\scriptsize
\[
\mi{CFC}(PN) = \sum_{t \in T \wedge (\abs{\preset{t}} > 1 \vee \abs{\postset{t}} > 1)} 1 + \sum_{p \in P \wedge (\abs{\preset{p}} > 1 \vee \abs{\postset{p}} > 1)} \abs{\postset{p}}
\]}
\end{itemize}

Let $(\syallacts, \chld, \alpha, \beta)$ be a hierarchical model returned. Let $q \in \{\qfit, \qprec, \mi{F1}, \mi{Ge}, \mi{CFC}, \mi{Size} \}$ be any quality measure that takes a model $M$ and/or a log $L$ as inputs and returns the quality value of the model. We calculate and report the average quality measure $\overline{q}$ as follows : 
$\overline{q}(\syallacts, \chld, \alpha, \beta) = \mi{avg}_{ a \in \syallacts \wedge \chld(a) \neq \emptyset }  q(\beta(a), \alpha(a)) $. For example, $\overline{\qfit}(\syallacts, \chld, \alpha, \beta)$ is calculated as the average value of the individual fitness values of each subprocess in the hierarchical model.

The computation of fitness, precision, and generalization uses a state-of-the-art technique known as alignment~\cite{adriansyah2011conformance}\footnote{A 60 minute time-out limit is set for computing the alignment of each model. When the particular quality measurement could not be obtained due to either syntactical or behavioral issues in the discovered model or a timeout when computing the quality measures, we record it using the ``$-$'' symbol. }.
As mentioned, we also list the quality values reported by Augusto et al. 
using ``\disflat''~\cite{DBLP:journals/tkde/AugustoCDRMMMS19}. Since we were unable to compute all qualities of the flat models, and for the sake of consistency, we mainly discuss our results with respect to the results of ``\disflat''. 

On the seven starred data sets, there are two subprocesses for \aamr-SM, one 
for \aamdk-SM, and
none for \aamdk-IMf and \aamr-IMf, where the returned alignments were indicated unreliable. As a result, the fitness of these models are -1, and subsequently, the precision and generalization cannot be computed. The results of these models are excluded when computing the average quality scores. 

\begin{table}[tbp]
	\caption{Evaluation results of models for the seven logs.}
	\begin{center}
			\resizebox{.48\textwidth}{!}{
			\begin{tabular}{lllrrrrrrrr}
				\toprule
				Data  & CAlg  & DAlg  & $\overline{Fi}$    & $\overline{Pr}$    & $\overline{F1}$    & $\overline{Ge}$    & $\overline{CFC}$    & $\overline{Size}$  & $\#$SPs & $\overline{\#Act}$  \\

\midrule
\multirow{8}[4]{*}{\rotatebox[origin=c]{90}{BPIC12}} & \aamflat    & \multirow{4}[2]{*}{IMf} & -     & -     & -     & -     & 66    & 115   & 1     & 24 \\
      & \disflat    &       & \textbf{0.98} & 0.50  & 0.66  & 0.66  & 37    & 59    & 1     & - \\
      & \aamdk &       & 0.96  & \textbf{0.78} & \textbf{0.86} & 0.85  & \textbf{20} & 36    & 4     & 8 \\
      & \aamr &       & 0.97  & \textbf{0.78} & \textbf{0.86} & 0.88  & 20    & \textbf{35} & 4     & 8 \\
\cmidrule{2-11}      & \aamflat    & \multirow{4}[2]{*}{SM} & \textbf{0.97} & 0.55  & 0.70  & 0.69  & 53    & 89    & 1     & 24 \\
      & \disflat    &       & 0.75  & 0.76  & 0.75  & 0.76  & 32    & 53    & 1     & - \\
      & \aamdk &       & 0.89  & \textbf{0.94} & \textbf{0.91} & \textbf{0.91} & \textbf{10} & \textbf{22} & 4     & 8 \\
      & \aamr &       & 0.92  & 0.90  & 0.90  & 0.90  & 14    & 28    & 3     & 8 \\
\midrule
\multirow{8}[4]{*}{\rotatebox[origin=c]{90}{BPIC17f}} & \aamflat    & \multirow{4}[2]{*}{IMf} & \textbf{0.98} & 0.57  & 0.72  & 0.73  & 20    & 51    & 1     & 18 \\
      & \disflat    &       & \textbf{0.98} & 0.70  & 0.82  & 0.82  & 20    & 35    & 1     & - \\
      & \aamdk &       & 0.97  & \textbf{0.98} & \textbf{0.97} & \textbf{0.97} & \textbf{6} & \textbf{17} & 4     & 6 \\
      & \aamr &       & \textbf{0.98} & 0.90  & 0.94  & 0.92  & 9     & 22    & 3     & 7 \\
\cmidrule{2-11}      & \aamflat    & \multirow{4}[2]{*}{SM} & \textbf{0.98} & 0.63  & 0.76  & 0.76  & 23    & 47    & 1     & 18 \\
      & \disflat    &       & 0.95  & 0.85  & 0.90  & 0.90  & 17    & 32    & 1     & - \\
      & \aamdk &       & 0.96  & \textbf{0.98} & \textbf{0.97} & \textbf{0.97} & \textbf{6} & \textbf{18} & 4     & 6 \\
      & \aamr &       & \textbf{0.98} & 0.92  & 0.95  & 0.95  & 8     & 20    & 3     & 7 \\
\midrule
\multirow{8}[4]{*}{\rotatebox[origin=c]{90}{BPIC15\_1f}} & \aamflat    & \multirow{4}[2]{*}{IMf} & 0.96  & 0.36  & 0.52  & 0.51  & 128   & 217   & 1     & 70 \\
      & \disflat    &       & 0.97  & 0.57  & 0.72  & 0.72  & 108   & 164   & 1     & - \\
      & \aamdk &       & \textbf{0.99} & \textbf{0.87} & \textbf{0.91} & \textbf{0.93} & \textbf{9} & \textbf{19} & 15    & 6 \\
      & \aamr &       & 0.97  & 0.84  & 0.88  & 0.88  & 16    & 29    & 9     & 10 \\
\cmidrule{2-11}      & \aamflat    & \multirow{4}[2]{*}{SM} & 0.93  & 0.87  & 0.90  & 0.90  & 64    & 152   & 1     & 70 \\
      & \disflat    &       & 0.90  & 0.88  & 0.89  & 0.89  & 43    & 110   & 1     & - \\
      & \aamdk &       & \textbf{0.96} & 0.98  & \textbf{0.97} & \textbf{0.97} & \textbf{4} & \textbf{14} & 15    & 6 \\
      & \aamr &       & 0.88  & \textbf{0.99} & 0.93  & 0.93  & 8     & 22    & 9     & 10 \\
\midrule
\multirow{8}[4]{*}{\rotatebox[origin=c]{90}{BPIC15\_2f}} & \aamflat    & \multirow{4}[2]{*}{IMf} & -     & -     & -     & -     & 183   & 313   & 1     & 82 \\
      & \disflat    &       & 0.93  & 0.56  & 0.70  & 0.70  & 123   & 193   & 1     & - \\
      & \aamdk &       & \textbf{0.97} & \textbf{0.91} & \textbf{0.93} & \textbf{0.93} & \textbf{7} & \textbf{16} & 21    & 5 \\
      & \aamr &       & 0.95  & 0.85  & 0.89  & 0.88  & 16    & 33    & 10    & 10 \\
\cmidrule{2-11}      & \aamflat    & \multirow{4}[2]{*}{SM} & 0.83  & 0.88  & 0.85  & 0.85  & 72    & 198   & 1     & 82 \\
      & \disflat    &       & 0.77  & 0.90  & 0.83  & 0.82  & 41    & 122   & 1     & - \\
      & \aamdk &       & \textbf{0.94} & \textbf{0.99} & \textbf{0.97} & \textbf{0.97} & \textbf{4} & \textbf{13} & 21    & 5 \\
      & \aamr &       & 0.83  & 0.96  & 0.89  & 0.89  & 11    & 26    & 10    & 10 \\
\midrule
\multirow{8}[4]{*}{\rotatebox[origin=c]{90}{BPIC15\_3f}} & \aamflat    & \multirow{4}[2]{*}{IMf} & -     & -     & -     & -     & 136   & 244   & 1     & 62 \\
      & \disflat    &       & 0.95  & 0.55  & 0.70  & 0.69  & 108   & 159   & 1     & - \\
      & \aamdk &       & \textbf{0.97} & \textbf{0.94} & \textbf{0.95} & \textbf{0.95} & \textbf{6} & \textbf{15} & 17    & 5 \\
      & \aamr &       & 0.94  & 0.87  & 0.90  & 0.89  & 16    & 33    & 8     & 10 \\
\cmidrule{2-11}      & \aamflat    & \multirow{4}[2]{*}{SM} & 0.81  & 0.92  & 0.86  & 0.87  & 50    & 135   & 1     & 62 \\
      & \disflat    &       & 0.78  & 0.94  & 0.85  & 0.85  & 29    & 90    & 1     & - \\
      & \aamdk &       & \textbf{0.95} & \textbf{0.99} & \textbf{0.97} & \textbf{0.97} & \textbf{3} & \textbf{12} & 17    & 5 \\
      & \aamr &       & 0.87  & 0.98  & 0.92  & 0.92  & 10    & 23    & 7     & 9 \\
\midrule
\multirow{8}[4]{*}{\rotatebox[origin=c]{90}{BPIC15\_4f}} & \aamflat    & \multirow{4}[2]{*}{IMf} & -     & -     & -     & -     & 121   & 242   & 1     & 65 \\
      & \disflat    &       & 0.96  & 0.58  & 0.72  & 0.72  & 111   & 162   & 1     & - \\
      & \aamdk &       & \textbf{0.98} & \textbf{0.96} & \textbf{0.97} & \textbf{0.97} & \textbf{5} & \textbf{13} & 21    & 4 \\
      & \aamr &       & 0.97  & 0.81  & 0.87  & 0.87  & 18    & 35    & 8     & 10 \\
\cmidrule{2-11}      & \aamflat    & \multirow{4}[2]{*}{SM} & 0.79  & 0.92  & 0.85  & 0.85  & 51    & 147   & 1     & 65 \\
      & \disflat    &       & 0.73  & 0.91  & 0.81  & 0.80  & 31    & 96    & 1     & - \\
      & \aamdk &       & \textbf{0.95} & \textbf{0.99} & \textbf{0.97} & \textbf{0.97} & \textbf{2} & \textbf{11} & 21    & 4 \\
      & \aamr &       & 0.83  & 0.98  & 0.90  & 0.89  & 8     & 24    & 8     & 10 \\
\midrule
\multirow{8}[4]{*}{\rotatebox[origin=c]{90}{BPIC15\_5f}} & \aamflat    & \multirow{4}[2]{*}{IMf} & -     & -     & -     & -     & 142   & 255   & 1     & 74 \\
      & \disflat    &       & 0.94  & 0.18  & 0.30  & 0.61  & 95    & 134   & 1     & - \\
      & \aamdk &       & \textbf{0.98} & \textbf{0.95} & \textbf{0.97} & \textbf{0.96} & \textbf{5} & \textbf{14} & 21    & 5 \\
      & \aamr &       & 0.98  & 0.80  & 0.87  & 0.87  & 19    & 34    & 9     & 10 \\
\cmidrule{2-11}      & \aamflat    & \multirow{4}[2]{*}{SM} & 0.85  & 0.91  & 0.88  & 0.88  & 54    & 163   & 1     & 74 \\
      & \disflat    &       & 0.79  & 0.94  & 0.86  & 0.85  & 30    & 102   & 1     & - \\
      & \aamdk &       & \textbf{0.96} & \textbf{1.00} & \textbf{0.98} & \textbf{0.98} & \textbf{2} & \textbf{11} & 20    & 5 \\
      & \aamr &       & 0.83  & 0.98  & 0.90  & 0.90  & 8     & 23    & 9     & 10 \\
\midrule
\end{tabular}%
	}
		\label{tab:result}%
	\end{center}
\end{table}%

\subsection{Results and Discussion}
\begin{figure*}[tb]
	\centerline{\includegraphics[width=\textwidth]{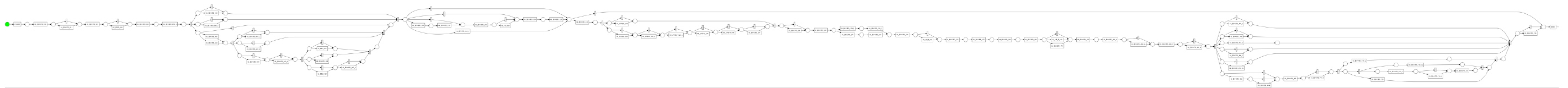}}
	\caption{The flat model discovered using \aamflat-IMf on the BPIC15\_1f log.}
	\label{fig:resBPIC151f}
\end{figure*}

\autoref{tab:result} summarises the results of our evaluation. For each data set (\columnn{Data}) and each of the two discovery algorithms (\columnn{DAlg}), we obtained four (hierarhical) models and report on the results. To also provide concrete examples, 
\autoref{fig:resBPIC151f}(a) shows the root process and three subprocesses obtained by applying \aamdk-IMf on the BPIC15\_1f log, while the other subprocesses are hidden (abstracted). By contrast, \autoref{fig:resBPIC151f}(b) shows the flat model by \aamflat-IMf on the same log.

\begin{figure*}[tb]
	\centerline{\includegraphics[width=\textwidth]{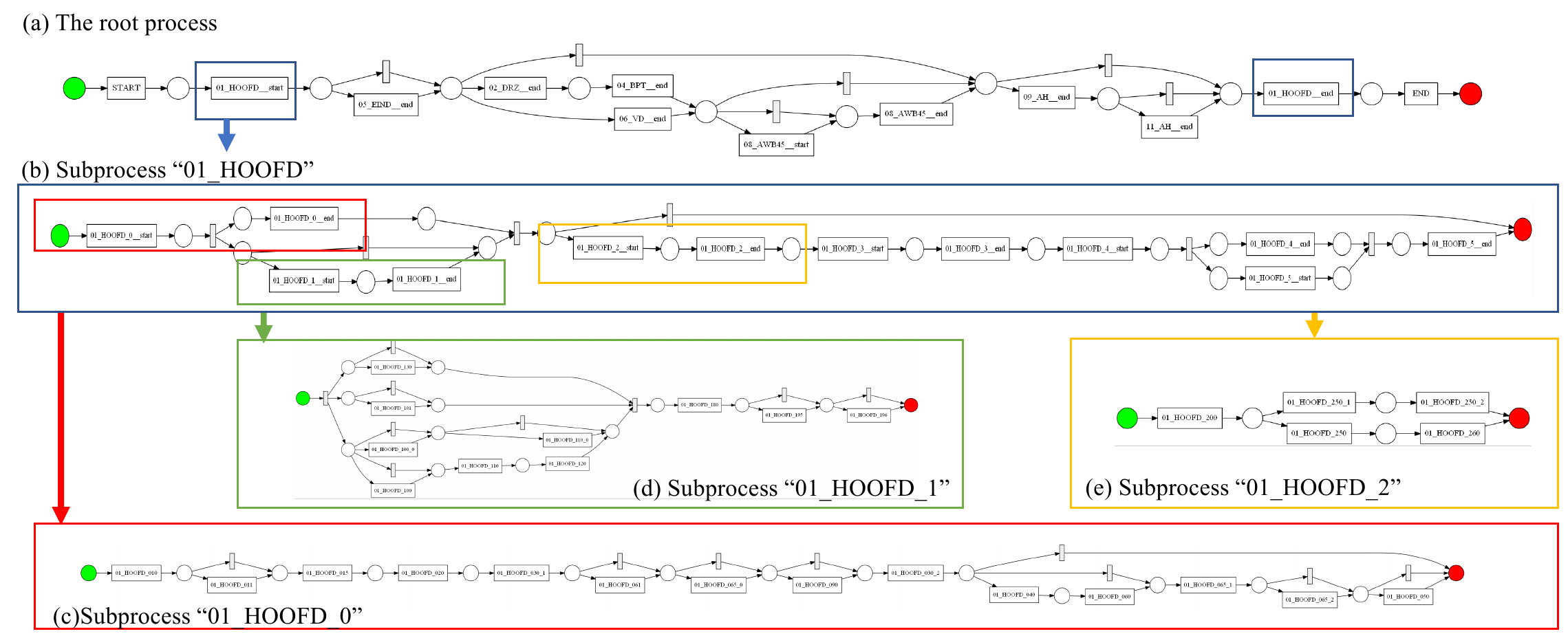}}
	\caption{The root process and three subprocesses discovered by \aamdk-IMf on the BPIC15\_1f log. }
	\label{fig:resBPIC151fsubs}
\end{figure*}




\textbf{Fitness.} 
Column \columnn{$\overline{Fi}$} (\autoref{tab:result}) reports on the average fitness of the models discovered. 
For both IMf and SM, in 5 of the seven logs, the hierarchical models returned by \aamdk achieved a better fitness score than \disflat. More specifically, there is on average an increase of 0.015 (2\%) in fitness, when compared the fitness of F* to \aamdk.  
Considering SM, the improvement in fitness is more significant: the average increase in fitness is 17\%, from 0.810 to 0.946. The moderate improvement in fitness for IMf is due to the fact that IMf tends to optimize fitness. As the average fitness of the seven models is already very high (0.959), there is little room for improvement. 

\textbf{Precision.} Column \columnn{$\overline{Pr}$} (\autoref{tab:result}) lists the average precision of the models. 
When comparing the average precision scores of the subprocesses (\aamdk) to the flat models (\disflat), \aamdk has achieved a considerable improvement in precision, especially for IMf. 
For all seven logs, the subprocesses obtained by both \aamdk-IMf and \aamdk-SM have achieved an average precision higher than the \disflat approaches: for \aamdk-IMf, the average precision is increased by 75.4\% (from 0.520 to 0.912);
for \aamdk-SM, there is also an 11.2\% increase in the precision scores (from 0.883 to 0.982). 

An explanation for such significant improvements in the average precision can be the following. As the subprocesses are relatively small and the sublogs do not contain interleaved behavior of other concurrent subprocesses, it allowed the discovery algorithm to discover rather sequential models of high precision, while maintaining the high fitness (see \autoref{fig:resBPIC151f} (b), (c) and (e)). When a subprocess itself contains much concurrent behavior (for example, see \autoref{fig:resBPIC151f} (d)), the highly concurrent, flexible behavior is then localized within this one subprocess, without affecting the models of other subprocesses. 

\textbf{F1 score.} 
For both IMf and SM and for all seven logs, the average F1 scores of the subprocesses of \aamdk outperform the ones of \disflat. This is mainly due to the improvement in both fitness (for SM) and precision (for IMf), which led to the increase in the harmonic mean of fitness and precision. On average \aamdk achieved an increase of 41.9\% (from 0.660 to 0.936) for IMf and 14.4\% for SM (from 0.842 to 0.962) in the F1 scores over the seven logs. 

\begin{figure}[tbp]
	\centerline{\includegraphics[width=0.5\textwidth]{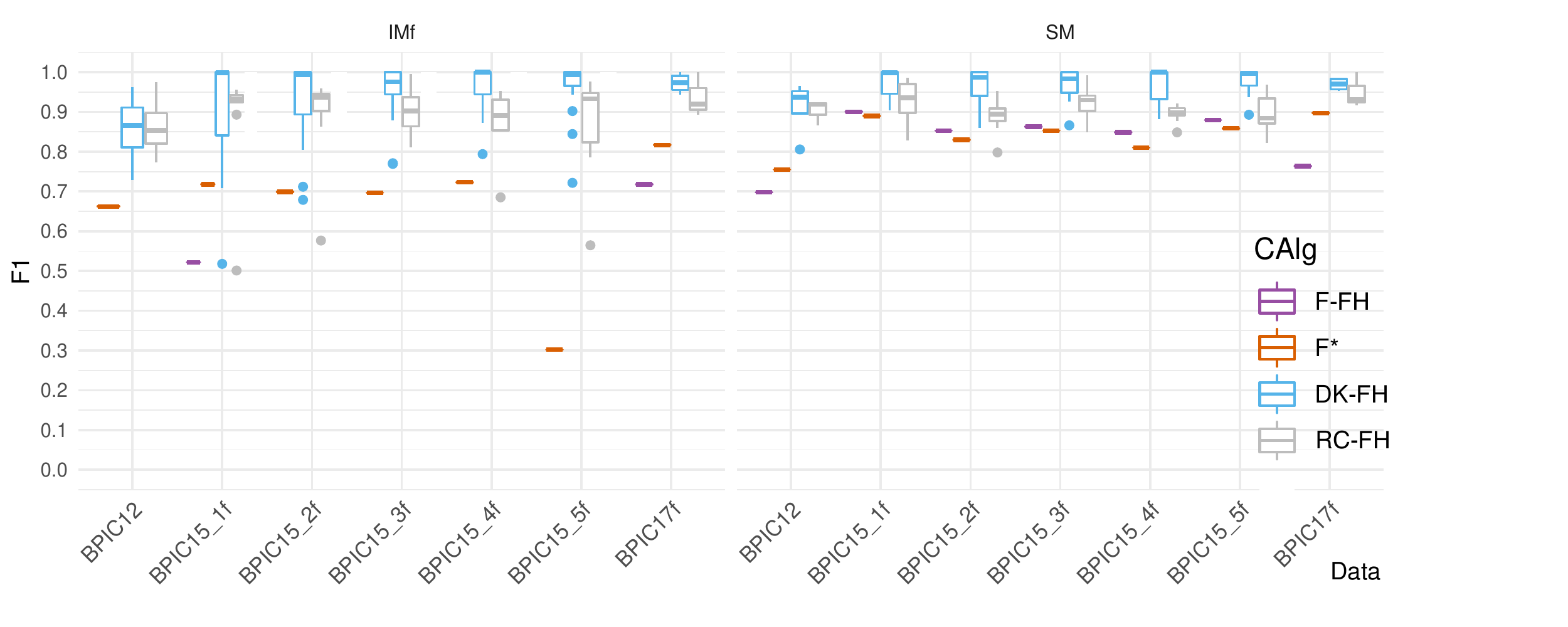}}
	\caption{Significantly higher F1-scores achieved by \aamdk (blue), followed by \aamr (grey), compared to the \aamflat (purple) and the results reported in \cite{DBLP:journals/tkde/AugustoCDRMMMS19} (purple), on the seven benchmark data sets (on the left IMf, and on the right SM).}
	\label{fig:resfit}
\end{figure}

\autoref{fig:resfit} shows the distribution of the F1-scores of the models of each approach in more detail (IMf on the left and SM on the right). Compared to the F1 scores of the models by \aamflat (see purple lines), \aamdk (blue boxplot and dots) always scores higher. Compared to the \disflat results, there are only three exceptional cases for IMf. 
We looked into these models of IMf and found that many activities are put in parallel. Thus, fitness was very high (e.g., 0.972) but precision was very low (e.g., 0.335).   
Interestingly, for the exact same subprocess, SM was able to discover a much more sequential model. This model still has a high fitness (0.958) but a much higher precision (1.00) and also a high generalization (0.979). This can be an interesting future work: since \aamdk allows for the use of any discovery algorithm, one can design an algorithm that selects the best algorithm (model) for each subprocess to further increase the quality of the hierarchical models.

\textbf{Generalization.} Column $\overline{Ge}$ (\autoref{tab:result}) reports on the generalization scores (using 3-fold cross validation). 
In all seven logs, for both IMf and SM, there is overall an increase of 23.2\% (from 0.771 to 0.950) in the generalization scores (33.3\% for IMf and 14.8\% for SM), when comparing \aamdk to \disflat.

\textbf{Complexity.}
Since the subprocesses are much smaller than the flat model, the average complexity scores of subprocesses of \aamdk are, as expected, significantly lower than \aamflat or \disflat. For \aamdk-IMf, the CFC is decreased by 90.6\% (from 86.0 to 8.1), and 86.3\% for \aamdk-SM (from 31.9 to 4.4), when compared to the CFC scores of the models of \disflat on the seven logs. The improvement is even more significant when it is compared to the models by \aamflat. 

\begin{figure}[tb]
	\centerline{\includegraphics[width=0.5\textwidth]{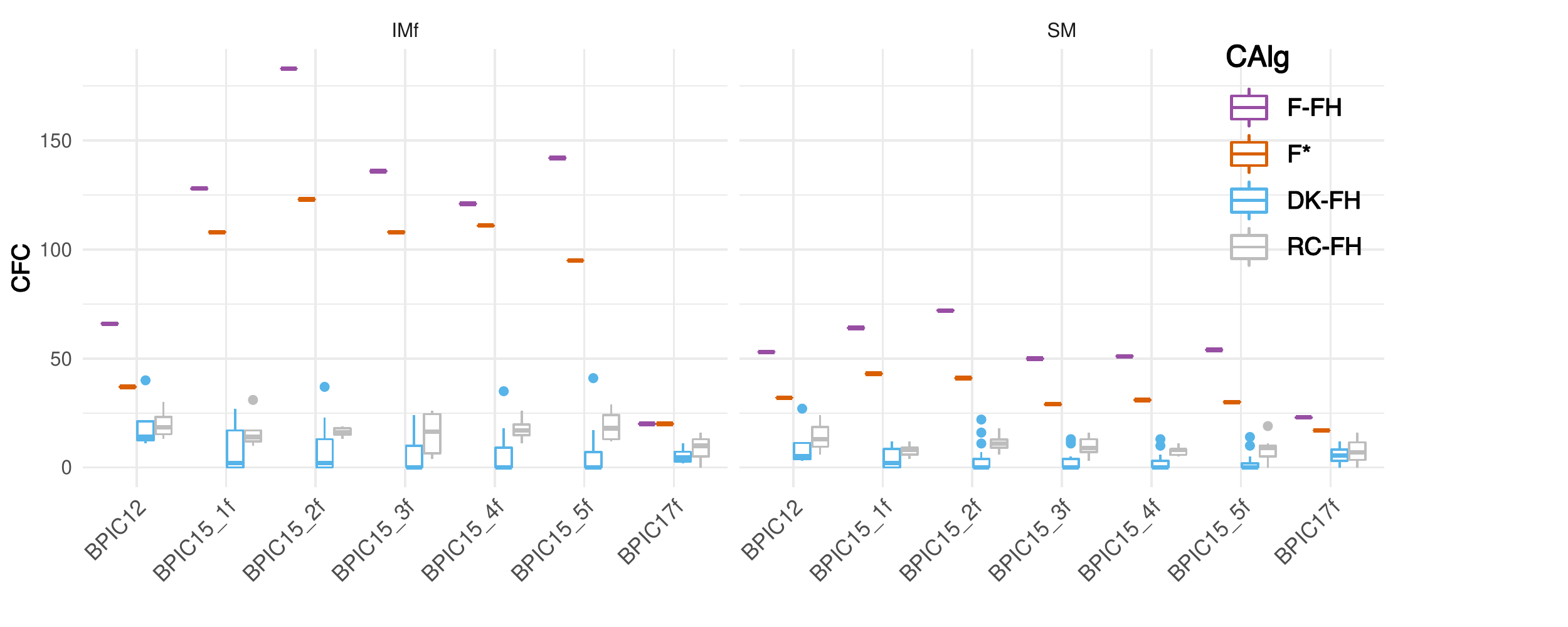}}
	\caption{A significant decrease in the complexity (CFC) achieved by the \aamdk (blue), compared to the \aamflat (F) returned (purple lines) for the 16 data sets. }
	\label{fig:resPrec}
\end{figure}

\autoref{fig:resPrec} shows the distribution of CFC of the submodels by two approaches in more detail: \aamdk (blue boxplots) and \aamflat (purple lines). It shows the significant decrease in CFC when using \aamdk. 
%

\textbf{Discussion.}
Overall, the results on the real-life logs have shown that 
the quality of the submodels returned by the \aamdk approach are higher than the one returned by either the random clustering approachs or the flat discovery algorithms. For example, compared to \disflat, \aamdk achieved an increase of 0.07 in fitness, 0.25 in precision, and 0.18 in generalization, on average. As expected, the \aamdk outperforms the \aamr.

Interestingly, the random activity clustering approach (\aamr) also shows relatively consistent improvements in the qualities of obtained submodels, compared to the flat tree approach and the flat discovery (F*), as listed in~\autoref{tab:result} and shown in~\autoref{fig:resfit}. This may suggest that discovering hierarchical models may have a certain beneficial factor in terms of model qualities, comparing to discovering a complex, flat model. Additional experiments are needed to validate this observation. However, if this is indeed the case, this may allow future process discovery algorithms to focus on small processes (say less than 10-20 activities), while designing other algorithms for clustering the activities and discovering the process hierarchy. 

A related, interesting observation is that for some subprocesses and their sublogs, SM was able to discover better models than IMf, and in other cases, IMf was able to find better models than SM. Since \aam allows for the use of any discovery algorithm, this suggests that one can build in a selection strategy of the best subprocesses to further increase the quality of the hierarchical models, as future work. 

We also observed that it is rather trivial to flatten the full model or certain subprocesses of interest, while allowing abstracting the others, see~\autoref{fig:resBPIC151fsubs}. For example, in~\autoref{fig:exampleApproach}(c) and (d), the logs $L2$ and $L3$ respectively show abstracting one or two subprocesses. Applying a discovery algorithm on $L2$ returns a model where subprocess $\actlabel{C}$ is abstracted; and for $L3$, a model is returned where subprocesses $\actlabel{C}$ and $\actlabel{L}$ are abstracted, while the detailed activities of subprocess $\actlabel{S}$ are retained.

\section{Related Work}
\label{sec:relatedwork}
In this section, we discuss related works regarding hierarchical process discovery and event abstraction, specifically.

Following the unsupervised strategy, Bose and van der Aalst~\cite{DBLP:conf/bpm/BoseA09} are one of the first who propose to detect reoccurring consecutive activity sequences as subprocesses.  
This approach treats concurrent subprocesses as running sequentially and thus does not assume true concurrent/interleaving subsprocesses. 
Later, Tax et al.~\cite{DBLP:journals/is/TaxDSAN18} propose an approach to generate frequent subprocesses, also known as \emph{local process models}. This approach enables detecting interleaving subprocess in an unsupervised manner, which is used to create abstracted logs. 


Using the supervised strategy, the State Chart Miner (SCM)~\cite{DBLP:conf/wcre/LeemansAB18} extends Inductive Miner (IM) and uses information in the activity labels to create a hierarchy to discover hierarchical models. SCM also focuses on sequential subprocesses, where non-consecutive events of the same instance are cut into separate traces. This assumption leads to the discovery of models that are overly segmented and fail to capture concurrent behavior, as shown in~\autoref{fig:rwSCMvsAAMbpi2012highscm}.
Furthermore, SCM can only use IM for discovering subprocesses. 
Mannhardt et al.~\cite{DBLP:conf/bpm/MannhardtLRAT16} require users to define complete behavioral models of subprocesses and their relations to compute abstracted logs. 
This prerequisite of specifying the full behavior of low-level processes put much burden on the users. Moreover, it uses the alignment technique to compute high-level logs which is computationally very expensive and can be nondeterministic. 


%
%
%
%


%
Other approaches assume additional attributes to indicate the hierarchical information. 
Using a relational database as input, Conforti et al.~\cite{DBLP:journals/is/ConfortiDGR16} assume that each event has a set of primary and foreign keys that can be used as the subprocess instance identifier, in order to determine subprocesses and multi-instances. However, such event attributes may not be common in most of the event logs.
%
As an alternative, Wang et al.~\cite{DBLP:conf/otm/WangWYSW15} assume that the events contain start and complete timestamps and have explicit information of the follow-up events (i.e., the next intended activity, which they called ``transfer'' attributes). 
Senderovich et al.~\cite{senderovich2015discovery} propose the use of patient schedules in a hospital as an approximation to the actual life-cycle of a visit. 
These approaches cannot be applied in our case, since we do not find these attributes (such as explicit causal-relations between events, the start and complete timestamps, or the scheduling of activities) in our logs. 

A very recent literature review conducted by van Zelst et al. \cite{vanzelst2020} provides an extensive overview and taxonomy for classifying different event abstraction approaches. 
In \cite{vanzelst2020}, Zelst et al. studied 21 articles in depth and compared them among seven different dimensions. One dimension is particularly important to distinguish our approaches, namely the supervision strategy. None of the 21 methods enable both supervised and unsupervised, whereas we have shown that we can used both supervised (domain knowledge) and unsupervised (random clustering) to discover an activity tree for event abstraction. 
Our evaluation has shown \aam to be flexible and applicable in practice.  It is worthy to mention that our approach does assume that for each case, each subprocess is executed at most once (i.e., single-instance), while a subprocess can contain loops. As future work, we can extend the algorithm to include, in addition to the abstraction and projection functions, the multi-instance detection and segmentation techniques to handle multi-instance subprocesses. 

\section{Conclusion}
\label{sec:conclusion}
In this work, we investigated the hierarchical process discovery problem. We presented \aam, a general approach that supports flexible ways to compute process hierarchy using the notion of \emph{activity tree}. We demonstrate this flexibility by proposing three methods for computing the hierarchy, which vary from fully supervised, using domain knowledge, to fully automated, using a random approach. We investigated the quality of hierarchical models discovered using these different methods. 
The empirical evidence, using seven real-life logs, demonstrates that the one using supervised approach outperforms the one using random clustering in terms of the four quality dimensions. But both methods outperform the flat model approaches, which clearly demonstrates the strengths of the \myminer~approach.
For future work, we plan to investigate different algorithms for computing activity trees to further improve the quality of hierarchical models. Also, the concept of activity trees can be extended to data-aware or context-aware activity trees. For example, an activity node can be associate with a data contraint (context), so that the events labeled with the same activity but have different data attributes (contexts) can be abstracted into different subprocesses. 

\section*{Acknowledgment}
This research was supported by the NWO TACTICS project (628.011.004) and Lunet Zorg in the Netherlands. We would also like to thank the experts from the VUMC for their extremely valuable assistance and feedback in the evaluation.


%
%
%
%

\bibliographystyle{IEEEtran}
\bibliography{reference}



\end{document}